\documentclass[conference,10pt]{IEEEtran}
\usepackage{epsfig,epsf,rotating,setspace,latexsym,amsmath,amssymb,amsfonts,bm,theorem,subfigure,epstopdf}
\usepackage{cite,authblk}
\usepackage{bbm}
\usepackage{algorithm}
\usepackage[noend]{algpseudocode}
\usepackage{color}
\usepackage{mathtools}
\usepackage{soul}

\algrenewcommand\algorithmicforall{\textbf{foreach}}
\algrenewcommand\algorithmicindent{.8em}

\newtheorem{lemma}{Lemma}

\newenvironment{Proof}[1]{\medskip\par\noindent{\bf Proof:\,}\,#1}{{\mbox{\,$\blacksquare$}\par}}

\IEEEoverridecommandlockouts
\allowdisplaybreaks

\begin{document}
 
\title{Age of Information With Non-Poisson Updates in Cache-Updating Networks}
 

 \author{Priyanka Kaswan \qquad Sennur Ulukus\\
         \normalsize Department of Electrical and Computer Engineering\\
         \normalsize University of Maryland, College Park, MD 20742\\
         \normalsize  \emph{pkaswan@umd.edu} \qquad \emph{ulukus@umd.edu}}
 
\maketitle

\begin{abstract}
We study age of information in multi-hop multi-cast cache-enabled networks where the inter-update times on the links are not necessarily exponentially distributed. We focus on the set of non-arithmetic distributions for inter-update times, which includes continuous probability distributions as a subset. We first characterize instantaneous age of information at each node for arbitrary networks. We then explicate the recursive equations for instantaneous age of information in multi-hop networks and derive closed form expressions for expected age of information at an end-user. We show that expected age in multi-hop networks exhibits an additive structure. Further, we show that the expected age at each user is directly proportional to the variance of inter-update times at all links between a user and the source. We expect the analysis in this work to help alleviate the over-dependence on Poisson processes for future work in age of information.
\end{abstract}

\section{Introduction}\label{sect:introduction}
Emerging time-sensitive applications with progressively dynamic data coupled with storage capacity becoming increasingly cheap has prompted several works in the study of age of information in cache-updating systems. Such network models have practical applications in settings such as V2V communications in vehicular environments, and content distribution networks. Given the generation time of the freshest packet at time $t$ at a node as $u(t)$, the age of information at this node is defined as $t-u(t)$, thereby capturing the essence of freshness of information at the node through its age. 

\begin{figure}[t]
\centerline{\includegraphics[width=0.65\linewidth]{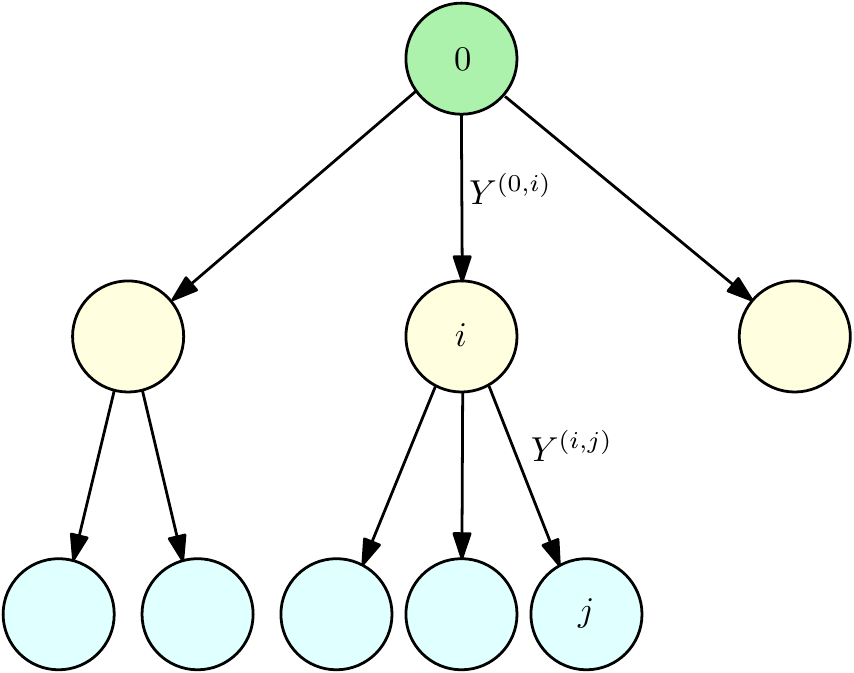}}
\vspace*{-0.1cm}
\caption{Multi-hop multi-cast tree network.}
\label{fig:tree network}
\vspace*{-0.4cm}
\end{figure}

In this work, we study age of information in a multi-hop multi-cast network with cache-aided nodes as shown in Fig.~\ref{fig:tree network}. We assume that the updates on each link are sent according to an ordinary renewal process which is not necessarily a Poisson process, and all nodes forward updates independently of other nodes. As a result, the evolution of age in the network is defined by a superposition of multiple independent ordinary renewal processes. As proved in \cite{samuels1974superpos}, the superposition of two ordinary renewal processes is an ordinary renewal process only if all processes are Poisson processes. 

Since the exponential distribution (or geometric distribution) is the only continuous (or discrete) probability distribution with memoryless property, most prior works studying timely information dissemination in networks heavily rely on these distributions. Here, it is worth noting how networks with nodes as queues \cite{yates18preempt, Kam18, Yates18parallel} differ from networks with nodes as caches. When a node has a queue buffer, all three random variables corresponding to, the arrival of a packet, waiting time of a packet in the queue and the service time of a packet, operate in a sequential order. In contrast, in cache-updating networks, each node requests packets from other nodes according to a renewal process, independently of how packets arrive at other nodes. Hence, simultaneous operation of renewal processes is distinctive to cache-updating networks. Nevertheless, when the random variables involved are memoryless, the age processes in cache-updating networks and preemptive-queuing networks are similar. This can be seen in the case of linear or tree networks by comparing the results of \cite{yates18preempt} and \cite{baturalp21clustergossip} for memoryless exponential random variables, and \cite{talak17aoimultihop} and \cite{selen13aoigossip} for memoryless geometric random variables. 

\begin{figure}[t]
 	\begin{center}
 	\subfigure[]
    {\includegraphics[width=0.35\linewidth]{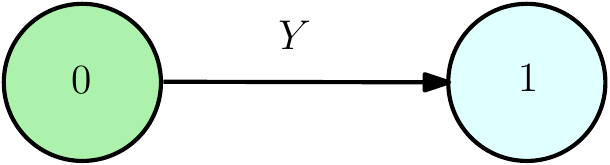}}
    \quad
 	\subfigure[]{\includegraphics[width=0.50\linewidth]{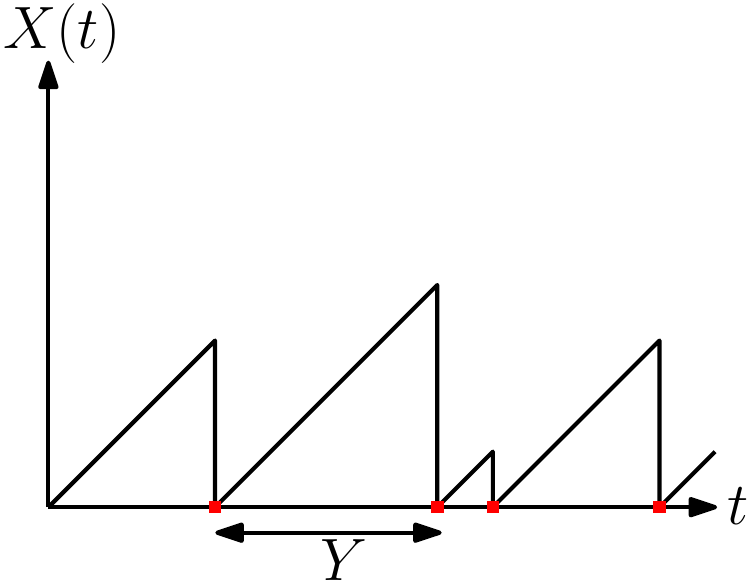}}
 	\end{center}
 	\vspace{-0.2cm}
 	\caption{(a) One-hop model, consisting of source (node $0$) and user (node $1$). (b) Red dots show times instants of packet arrival at user from source.}
 	\label{fig:one_hop_model}
 	\vspace{-0.4cm}
\end{figure}

However, when the inter-update times follow general distributions which are not exponential, it becomes difficult to calibrate the different renewal processes with respect to each other. To understand this challenge, let us first consider the single-hop system model of Fig.~\ref{fig:one_hop_model}(a) which offers the convenience of a single renewal process, and hence, circumvents the need to deal with superposition of arrival processes. In this model, the user downloads packets from the source according to a renewal process where inter-update times $Y_i$ are positive i.i.d.~random variables with non-arithmetic distribution $F$. In this respect, a distribution $F$ is called arithmetic (or periodic) if it is piecewise constant and its points of increase are contained in a set $\{0,d,2d, \ldots\}$ with the largest such $d>0$ being the span of such distribution. When $F$ is not arithmetic, it is called non-arithmetic, e.g., a distribution with a continuous part \cite{serfozo09}.

In the above model, the age of information $X(t)$ at the user, shown in Fig.~\ref{fig:one_hop_model}(b), is determined by the time elapsed since the last renewal. Since $X(t)$ at time $t$ depends only on location of $t$ within the inter-renewal interval containing $t$, $X(t)$ qualifies as a renewal reward process \cite{gallager11}, and its ensemble-average in limit of large $t$ is the same as its time average which, with probablity $1$, provided $\mathbb{E}[Y^2]<\infty$, is given by
\begin{align}\label{eqn:onehop_limit_expec}
    \lim_{t \to \infty}\mathbb{E}[X(t)]=\lim_{t \to \infty}\frac{\int_{\tau=0}^{t}X(\tau)d\tau}{t}= \frac{\mathbb{E}[Y^2]}{2\mathbb{E}[Y]}
\end{align}
where $Y$ is a typical random variable with distribution $F$. 

\begin{figure}[t]
 	\begin{center}
 	\subfigure[]
    {\includegraphics[width=0.6\linewidth]{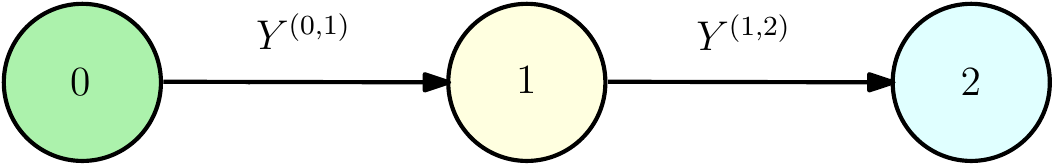}}\\
 	\subfigure[]{\includegraphics[width=0.9\linewidth]{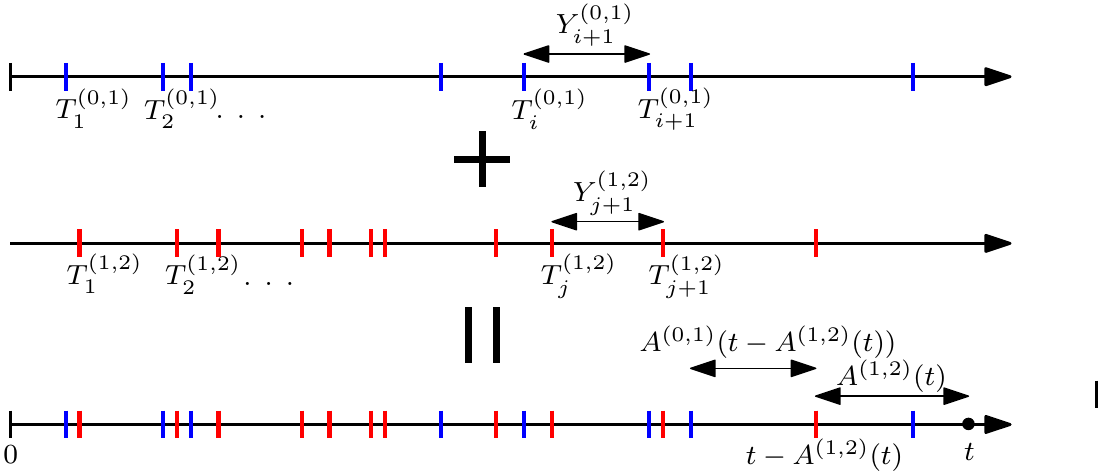}}
 	\end{center}
 	\vspace{-0.3cm}
 	\caption{(a) Two-hop model. (b) Superposition of two renewal processes, $N^{(0,1)}(t)$ and $N^{(0,1)}(t)$.}
 	\label{fig:two_hop_model}
 	\vspace{-0.55cm}
 \end{figure}

We next add a second link to this model which results in the two-hop network of Fig.~\ref{fig:two_hop_model}(a), where updates arrive at node $k$ from node $k-1$ at times $T^{(k-1,k)}_i$ according to renewal process $k$, as shown in Fig.~\ref{fig:two_hop_model}(b). If we consider a typical inter-renewal interval $[T^{(0,1)}_i,T^{(0,1)}_{i+1}]$, then in the absence of memoryless inter-update times, the distribution of packet arrival instant at node $2$ in this interval is dependent on when the last packet arrived at node $2$ in previous inter-renewal intervals, which prevents us from characterizing the age process in this interval independently of the past. Similarly, if we were to consider an inter-renewal interval $[T^{(1,2)}_j,T^{(1,2)}_{j+1}]$, then the user age at the beginning or end of this interval depends on arrivals under renewal process $1$ in prior intervals. Therefore, the age evolutions in different time intervals are correlated throughout the timeline, and an interesting question to ask here is, whether it is possible to somehow characterize the ensemble average of age at user $2$ in the regime of large $t$.

The works that are most relevant to our paper are as follows: In \cite{Yates17sqrt}, age of information is studied in a single-hop network with $N$ files having varying degrees of popularity. In \cite{bastopcu20_google}, freshness of a citation index is studied for $n$ researchers with Poisson citation arrivals. \cite{baturalp21clustergossip} and \cite{bastopcu2020LineNetwork} study multi-hop networks with Poisson arrivals for version age of information metric and binary freshness metric, respectively. \cite{kaswan_isit2021} focuses on two-hop parallel relay networks with Poisson arrivals. \cite{Yates21gossip_traditional} and \cite{Yates21gossip} have initiated the study of age of information and version age of information, respectively, in arbitrary networks with Poisson arrivals, using the stochastic hybrid system (SHS) framework \cite{Yates21gossip_traditional, Yates21gossip, Bastopcu21gossip, baturalp21clustergossip, maatouk22, kaswan22jammerring, kaswan22timestomping, mitra_allerton2022}. The particular convenience of Poisson updates in SHS stems from the fact that exponential distribution is the only distribution with constant hazard rate, which simplifies expressions involving expectations of products of test functions and transition intensities or hazard rates. 

In this paper, we first attempt to characterize age of information in cache-updating systems for non-Poisson renewal processes. Though getting expressions for the long-term expected age proves difficult for general networks, we provide a closed form expression for the expected age in multi-cast networks which exhibit a tree topology as shown in Fig.~\ref{fig:tree network}. Further, we observe from (\ref{eqn:onehop_limit_expec}) that for a fixed mean arrival time $\mathbb{E}[Y]$, the age of information in a single-hop network increases with an increase in the variance $var[Y]$, since $\mathbb{E}[Y^2]=var[Y]+\mathbb{E}[Y]^2$. We show that the age at the end-user in a multi-hop network also exhibits a similar relationship with the variance of the inter-arrival distribution of the user node, which gives all nodes an opportunity to lower their age, independent of the dynamics of the network. Finally, we present simulation results to support our findings.

\section{Model and Notations}\label{sect:model_notations}

Packets are assumed to arrive from node $i$ at node $j$ on link $(i,j)$ according to  $N^{(i,j)}(t)$ renewal process, and the finite random times $0\leq T^{(i,j)}_1 \leq T^{(i,j)}_2 \leq \ldots$ denote the corresponding renewal times, such that, inter-arrival times $Y^{(i,j)}_n=T^{(i,j)}_n- T^{(i,j)}_{n-1}$ are i.i.d.~with common distribution $F^{i,j}$. Given $N^{(i,j)}(t)=\max\{n:T^{(i,j)}_n\leq t\}$, the regenerative process $A^{(i,j)}(t)=t-T^{(i,j)}_{N^{(i,j)}(t)}$ denotes the corresponding backward recurrence time (or current life or age of renewal process) at $t$, which is the time since the last renewal prior to $t$. Note that $0 \leq A^{(i,j)}(t) \leq t$, which will be repeatedly used later. For more details, please see references \cite{serfozo09, gallager11}.

Consider a typical node $j$ in an arbitrary network of nodes and let $S_j$ denote the set of nodes from which packets arrive at node $j$. The most recent packet from node $i \in  S_j$ arrives at node $j$ before time $t$ at time instant $t-A^{(i,j)}(t)=T^{(i,j)}_{N^{(i,j)}(t)}$, at which point, node $j$ compares the generation times of the arriving packet with the packet present at its cache, and discards the staler packet in favor of the fresher packet. 

Let $X_j(t)$ denote the instantaneous age of information at node $j$ at time $t$. Then, $X_j(t)$ can be written as
\begin{align}\label{eqn:instantaneous_age_general_networks}
    X_j(t)=\sum_{i\in S_j}\Big[\prod_{k\in S_j\backslash\{i\}}&\chi_{\{A^{(i,j)}(t)< A^{(k,j)}(t)\}}\Big] \nonumber \\
    \times\Big[\min\big\{X_i&(t-A^{(i,j)}(t)),X_j(t-A^{(i,j)}(t)\big\} \nonumber \\
    &+A^{(i,j)}(t)\Big]
\end{align}
where $\chi_{\mathcal{A}}$ represents the indicator random variable for the measurable set $\mathcal{A}$ and $X_j(0)=0$. Since the source always has the latest packet, $X_0(t)=0$ at all times.\footnote{(\ref{eqn:instantaneous_age_general_networks}) holds true for the case where the distributions of inter-update times do not have atom points. When the distributions have atom points, packets from different nodes might arrive at node $j$ at the same time with non-zero probability. This situation can be remedied by choosing a priority order for different incoming links, which would change some of the ``$<$'' to ``$\leq$'' in the indicator variable in (\ref{eqn:instantaneous_age_general_networks}).}

In (\ref{eqn:instantaneous_age_general_networks}), $\prod_{k\in S_j\backslash\{i\}}\chi_{\{A^{(i,j)}(t)< A^{(k,j)}(t)\}}$ corresponds to the scenario when the last packet that arrived at node $j$ before time $t$ came from node $i$, which would be the case when the backward recurrence times of all other relevant renewal processes at time $t$ are larger than $A^{(i,j)}(t)$. The last term $A^{(i,j)}(t)$ in (\ref{eqn:instantaneous_age_general_networks}) comes from the fact that age grows at unit rate post the last packet arrival.

In the next step, $\min\{X_i(t-A^{(i,j)}(t)),X_j(t-A^{(i,j)}(t)\}$, which is a minimum over two age processes, can be further characterized in a manner similar to (\ref{eqn:instantaneous_age_general_networks}), and the corresponding expression will have terms of the form that involve taking a minimum over three age processes, for example $\min\{X_i(t'),X_j(t'),X_k(t')\}$ with $t'=t-A^{(i,j)}(t)- A^{(k,j)}(t-A^{(i,j)}(t))$. By recursively repeating this process we finally encounter the expression $\min\{X_1(t''),X_2(t''),\ldots,X_n(t'')\}$, $t''=t-\Delta(t)$, where $\Delta(t)$ represents a stochastic process whose exact expression depends on the network topology. Since source node is the only node external to the set of $n$ nodes, the last $\min$ expression can be completely defined in terms of backward recurrence times of the form $A^{(0,\ell)}(t''')$ such that $0\in S_{\ell}$. This recursive approach will become more clear for multi-cast networks in Section~\ref{sect:age_treenetworks}.

On first glance, this might give an impression that since all renewal processes and their associated recurrence times are independent processes, by reducing $X_j(t)$ to a function composed purely of backward recurrence times, one could conveniently compute the expectation of $X_j(t)$. However, note that, in the first and second steps of the recursion above, we encountered the term $A^{(k,j)}(t)$ in the product of indicator variables in (\ref{eqn:instantaneous_age_general_networks}), and the term $A^{(k,j)}(t-A^{(i,j)}(t))$ in the definition of $t'$. Though both terms correspond to the same renewal process $N^{(k,j)}(t)$, these backward recurrence times could be correlated through time which complicates analysis. 

However this complication does not arise if we assume that each node in the network has only one incoming link, as shown in the tree network of Fig.~\ref{fig:tree network}. This is because, $|S_j|=1$, and the product term in (\ref{eqn:instantaneous_age_general_networks}) vanishes. Additionally, since packets now arrive at node $j$ from a single preceding node $i$, $X_i(t)\leq X_j(t)$ for all $t$, which simplifies the $\min$ term as, 
\begin{align}
    \min\big\{X_i(t-A^{(i,j)}(t)),X_j&(t-A^{(i,j)}(t)\big\} \nonumber\\
    &=X_i(t-A^{(i,j)}(t))
\end{align}

In the next section, we derive a closed form expression for the long-term expected age $\lim_{t \to \infty}\mathbb{E}[X_j(t)]$ at each node $j$ in networks that have a tree topology.

\section{Age in Networks with Tree Structure}\label{sect:age_treenetworks}

\subsection{Two-Hop Network}
Consider the two-hop network in Fig.~\ref{fig:two_hop_model}(a), where we wish to determine the long-term expected age at node $2$. For this network, the instantaneous age $X_2(t)$ can be written as
\begin{align}\label{eqn:X_2_two_hop_recursive}
    X_2(t)=X_1(t-A^{(1,2)}(t))+A^{(1,2)}(t)
\end{align}
where $X_1(t-A^{(1,2)}(t))$ in turn can be expressed as
\begin{align}\label{eqn:X_1_two_hop_recursive}
    X_1(t-A^{(1,2)}(t))
    =&X_0(t-A^{(1,2)}(t)-A^{(0,1)}(t-A^{(1,2)}(t)))    \nonumber\\
    &+A^{(0,1)}(t-A^{(1,2)}(t))
\end{align}
Let us define 
\begin{align}
    \Delta_1(t)&=A^{(1,2)}(t)\label{eqn:delta1_2hop}\\
    \Delta_2(t)&=A^{(0,1)}(t-A^{(1,2)}(t))\label{eqn:delta2_2hop}
\end{align}
Substituting (\ref{eqn:X_1_two_hop_recursive}), (\ref{eqn:delta1_2hop}) and (\ref{eqn:delta2_2hop}) into (\ref{eqn:X_2_two_hop_recursive}), 
and using $X_0(t-\Delta_1(t)-\Delta_2(t))=0$, we obtain
\begin{align}\label{eqn:twohop_instantage}
    X_2(t)= \Delta_1(t)+\Delta_2(t)
\end{align}

\begin{lemma}\label{lemma:composition_2stoch_procs}
    Given independent stochastic processes $S_1(t)$ and $S_2(t)$ with $\sup_{t\geq 0}\left|\mathbb{E}[S_1(t)]\right|<\infty$ and $0\leq S_2(t)\leq t$, such that $\lim_{t \to \infty}\mathbb{E}[S_1(t)]$ and $\lim_{t \to \infty}\mathbb{E}[S_2(t)]$ exist, we have
    \begin{align}
        \lim_{t \to \infty}\mathbb{E}[S_1(t-S_2(t))]=\lim_{t \to \infty}\mathbb{E}[S_1(t)]
    \end{align}
\end{lemma}

\begin{Proof}
Let us define expectation functions
\begin{align}
    \mu_1(t)&=\mathbb{E}[S_1(t)]\\
    \mu_2(t)&=\mathbb{E}[S_2(t)]
\end{align}
such that $\mu_1(t)\to a_1$ and $\mu_2(t)\to a_2$, where $a_1$ and $a_2$ are some constants. Thus, for every $\epsilon_2>0$ there exists time $T_2$ such that for all $t> T_2$, we have $|\mu_2(t)-a_2|<\epsilon_2$. Choosing $\epsilon_2=a_2$, and using Markov inequality, we obtain
\begin{align}\label{eqn:Markov_ineq_upperbound}
    \mathbb{P}(S_2(t)\geq\sqrt{t})\leq\frac{\mu_2(t)}{\sqrt{t}}<\frac{2a_2}{\sqrt{t}}
\end{align}

Likewise, for every $\epsilon_1>0$, there exists $T_1$ such that for all $t> T_1$, we have $|\mu_1(t)-a_1|<\epsilon_1$. Since $S_1(t)$ and $S_2(t)$ are independent stochastic processes, for all $t\geq0$, we have 
\begin{align}
    \mathbb{E}[S_1(t-S_2(t))|S_2(t)]=\mu_1(t-S_2(t))
\end{align}
Therefore,
\begin{align}
    \Big|\mathbb{E}&\big[S_1(t-S_2(t))\big]-a_1 \Big|=\Big|\mathbb{E}\big[\mu_1(t-S_2(t))-a_1\big]\Big| \nonumber\\
    =&\Big|\mathbb{E}\left[ (\mu_1(t-S_2(t))-a_1)\left(\chi_{\{S_2(t)\geq\sqrt{t}\}} +\chi_{\{S_2(t)<\sqrt{t}\}}\right)\right]\Big| \nonumber\\
    \leq&\Big|\mathbb{E}\left[ \mu_1(t-S_2(t))\chi_{\{S_2(t)\geq \sqrt{t}\}}\right]\Big|+\Big|\mathbb{E}\left[a_1\chi_{\{S_2(t)\geq \sqrt{t}\}}\right]\Big|\nonumber\\
    &+\Big| \mathbb{E}\left[ (\mu_1(t-S_2(t))-a_1)\chi_{\{S_2(t)<\sqrt{t}\}}\right]\Big|\label{eqn:convg_expec_composition1_int}\\
    \leq&\mathbb{E}\left[\Big| \mu_1(t-S_2(t))\Big|\chi_{\{S_2(t)\geq \sqrt{t}\}}\right]+\mathbb{E}\left[a_1\chi_{\{S_2(t)\geq \sqrt{t}\}}\right]\nonumber\\
    &+\mathbb{E}\left[ \Big|\mu_1(t-S_2(t))-a_1\Big|\chi_{\{S_2(t)<\sqrt{t}\}}\right] \label{eqn:convg_expec_composition1}
\end{align}
where (\ref{eqn:convg_expec_composition1_int}) follows from the triangle inequality. 

If $t>T'=\frac{1+2T_1+\sqrt{1+4T_1}}{2}$, then when $S_2(t)<\sqrt{t}$, we have $t-S_2(t)> t-\sqrt{t}>T_1$ and hence $|\mu_1(t-S_2(t))-a_1|<\epsilon_1$. For all other cases, $|\mu_1(t-S_2(t))|\leq \sup_{t\geq 0}\left|\mathbb{E}[S_1(t)]\right|$. For $t>\max\{T',T_2\}$, using (\ref{eqn:Markov_ineq_upperbound}), (\ref{eqn:convg_expec_composition1}) gives
\begin{align}
    \Big|\mathbb{E}[S_1(t-S_2(t))]-a_1 \Big|<& \sup_{t\geq 0}\left|\mathbb{E}[S_1(t)]\right|\frac{2a_2}{\sqrt{t}} +\frac{2a_1a_2}{\sqrt{t}}+\epsilon_1
\end{align}
Since $\epsilon_1$ can be chosen to be arbitrarily small,
\begin{align}
    \lim_{t \to \infty}\Big|\mathbb{E}[S_1(t-S_2(t))]-a_1 \Big|=0
\end{align}
Thus, $\mathbb{E}[S_1(t-S_2(t))] \to a_1$, which completes the proof.
\end{Proof}

From (\ref{eqn:twohop_instantage}), $\mathbb{E}[X_2(t)]$ requires computing the terms $\mathbb{E}[\Delta_1(t)]$ and $\mathbb{E}[\Delta_2(t)]$. Since $A^{(1,2)}(t)$ evolves as in Fig.~\ref{fig:one_hop_model}(b), (\ref{eqn:onehop_limit_expec}) gives
\begin{align}\label{eqn:2hop_limit_delta1}
    \lim_{t \to \infty}\mathbb{E}[\Delta_1(t)]=\lim_{t \to \infty}\mathbb{E}[A^{(1,2)}(t)]=\frac{\mathbb{E}\left[\left(Y^{(1,2)}\right)^2\right]}{2\mathbb{E}\left[Y^{(1,2)}\right]}
\end{align}
Likewise, we have
\begin{align}
    \lim_{t \to \infty}\mathbb{E}[A^{(0,1)}(t)]=\frac{\mathbb{E}\left[\left(Y^{(0,1)}\right)^2\right]}{2\mathbb{E}\left[Y^{(0,1)}\right]}
\end{align}
Since the limit $\lim_{t \to \infty}\mathbb{E}[A^{(0,1)}(t)]$ exists, there exists $T$ such that for all $t>T$, $\mathbb{E}[A^{(0,1)}(t)]<\lim_{t \to \infty}\mathbb{E}[A^{(0,1)}(t)]+\epsilon$ for some $\epsilon>0$. Further, since $0\leq A^{(0,1)}(t) \leq t$ by definition, we have $\mathbb{E}[A^{(0,1)}(t)]<T$ for $t\leq T$. Hence, 
\begin{align}\label{eqn:2hop_supremum_01proof}
    \sup_{t\geq 0}\left|\mathbb{E}[A^{(0,1)}(t)]\right|\leq \max\big\{\lim_{t \to \infty}\mathbb{E}[A^{(0,1)}(t)]+\epsilon,T\big\}<\infty
\end{align}
Hence by Lemma~\ref{lemma:composition_2stoch_procs},
\begin{align}\label{eqn:2hop_limit_delta2}
    \lim_{t \to \infty}\mathbb{E}[\Delta_2(t)]=\lim_{t \to \infty}\mathbb{E}[A^{(0,1)}(t)]=\frac{\mathbb{E}\left[\left(Y^{(0,1)}\right)^2\right]}{2\mathbb{E}\left[Y^{(0,1)}\right]}
\end{align}
Thus, the long-term expected age at node $2$ is 
\begin{align}
    \lim_{t \to \infty}\mathbb{E}[X_2(t)]=\frac{\mathbb{E}\left[\left(Y^{(1,2)}\right)^2\right]}{2\mathbb{E}\left[Y^{(1,2)}\right]}+\frac{\mathbb{E}\left[\left(Y^{(0,1)}\right)^2\right]}{2\mathbb{E}\left[Y^{(0,1)}\right]}
\end{align}

\begin{figure}[t]
\centerline{\includegraphics[width=0.8\linewidth]{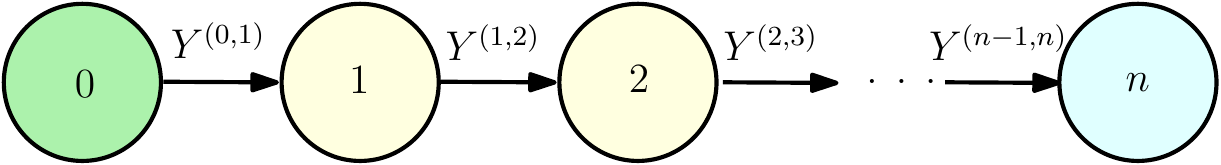}}
\caption{$n$-hop model.}
\label{fig:n_hop_model}
\vspace*{-0.4cm}
\end{figure}

Interestingly, the age at node $2$ is determined by independent contributions of links in the path from node $0$ to node $2$. In general, for tree networks, only the links involved in the path between source and an end-user are critical to the age dynamics of the end-user. In the next subsection, we therefore study $n$-hop linear networks.

\subsection{Multi-Hop Network}

\begin{figure}[t]
\centerline{\includegraphics[width=0.95\linewidth]{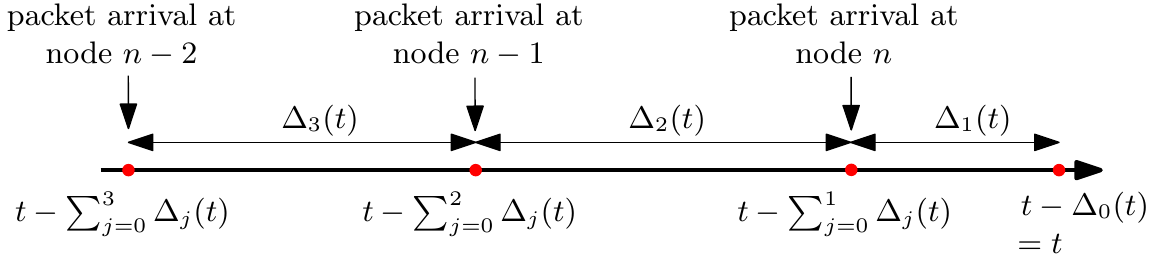}}
\caption{Time segments $\Delta_{i}(t)$ for $i\in \{ 1,\ldots,n\}$.}
\label{fig:deltas_on_timeline}
\vspace*{-0.4cm}
\end{figure}

Consider the $n$-hop network of Fig.~\ref{fig:n_hop_model} where we wish to determine the long-term expected age at node $n$ of the network. We define time segments $\Delta_{i}(t)$, $i\geq 1$ through the following recurrence equation
\begin{align}\label{eqn:multihop_deltarecursieve}
    \Delta_{i}(t)&=A^{(n-i,n-i+1)}(t-\sum_{j=0}^{i-1}\Delta_j(t))
\end{align}
with $\Delta_0(t)=0$, see Fig.~\ref{fig:deltas_on_timeline}. Note that $\Delta_{i}(t)$ is smaller than $t-\sum_{j=0}^{i-1}\Delta_j(t)$ by definition of $A^{(n-i,n-i+1)}(t)$. Similar to (\ref{eqn:X_2_two_hop_recursive}), the instantaneous age $X_n(t)$ at node $n$ can be written as
\begin{align}\label{eqn:multihop_Xnt_raw}
    X_n(t)=X_{n-1}(t-A^{(n-1,n)}(t))+A^{(n-1,n)}(t) 
\end{align}
This can be alternately represented as
\begin{align}
    X_n(t-\Delta_0(t))=X_{n-1}(t-\Delta_0(t)-\Delta_1(t)) +\Delta_1(t)
\end{align}

In the next step, $X_{n-1}(t-A^{(n-1,n)}(t))$ of (\ref{eqn:multihop_Xnt_raw}) will be again characterized in a similar manner and the full set of equations encountered in this recursive approach is of the form
\begin{align}\label{eqn:Xnit_recursiveeqn}
    X_{n-i}(t-\sum_{j=0}^{i}\Delta_j(t))=X_{n-i-1}(t-\sum_{j=0}^{i+1}\Delta_j(t)) +\Delta_{i+1}(t)
\end{align}
for $0\leq i\leq n-1$ with $X_0(t-\sum_{j=1}^{n}\Delta_j(t))=0$ as node $0$ represents the source node. Then, it follows from (\ref{eqn:Xnit_recursiveeqn}) that
\begin{align}\label{eqn:nhop_Xnt_sum_instantaneous}
    X_n(t)=\sum_{j=1}^{n}\Delta_j(t)
\end{align}

Similar to (\ref{eqn:2hop_limit_delta1}), we have 
\begin{align}
    \!\!\!\!\lim_{t \to \infty}\mathbb{E}[\Delta_1(t)]=\lim_{t \to \infty}\mathbb{E}[A^{(n-1,n)}(t)]=\frac{\mathbb{E}\left[\left(Y^{(n-1,n)}\right)^2\right]}{2\mathbb{E}\left[Y^{(n-1,n)}\right]}\!\!\!
\end{align}
Further, using the approach of (\ref{eqn:2hop_supremum_01proof}), we get $\sup_{t\geq 0}\left|\mathbb{E}[A^{(n-i,n-i+1)}(t)]\right|<\infty$. Since $\sum_{j=0}^{i-1}\Delta_j(t)\leq t$, we can prove $\lim_{t \to \infty}\mathbb{E}[\Delta_i(t)]=\lim_{t \to \infty}\mathbb{E}[A^{(n-i,n-i+1)}(t)]$ recursively for $i=2,3,\ldots,n$ from (\ref{eqn:multihop_deltarecursieve}) using Lemma~\ref{lemma:composition_2stoch_procs}.

Hence from (\ref{eqn:nhop_Xnt_sum_instantaneous}), we obtain
\begin{align}
    \lim_{t \to \infty}\mathbb{E}[X_n(t)]=&\sum_{j=1}^{n}\lim_{t \to \infty}\mathbb{E}[\Delta_j(t)]\\
    =&\sum_{j=1}^{n}\frac{\mathbb{E}\left[\left(Y^{(n-j,n-j+1)}\right)^2\right]}{2\mathbb{E}\left[Y^{(n-j,n-j+1)}\right]}\label{eqn:nhop_limit_expec_age_node_n}
\end{align}

In the special case when the renewal processes are Poisson, (\ref{eqn:nhop_limit_expec_age_node_n}) reduces to the sum of inverse of rates of the Poisson processes, which was observed in \cite{yates18preempt} for preemptive queues, and in \cite{baturalp21clustergossip} for cache-updating systems, using the SHS method.

Interestingly, the age at node $n$ depends on independent contributions of the intermediate links $(i,i+1)$, $0\leq i \leq n-1$ and is invariant to ordering of these links. Hence, each node can minimize its age by optimizing its individual packet request renewal process, irrespective of the statistical properties of other nodes and links in the network. 

Since the constant random variable has zero variance, for a fixed mean $\mathbb{E}[Y^{(n-j,n-j+1)}]$, (\ref{eqn:nhop_limit_expec_age_node_n}) hints that all nodes should request packets at constant time intervals to reduce variance. Here, we would like to point out that the constant distribution is an example of arithmetic distribution where the limit  $\lim_{t \to \infty}\mathbb{E}[X(t)]$ does not exist. This is because the ages at all nodes evolve as deterministic functions of time and the expectations do not stabilize at large $t$. Time average instead is the more reasonable choice of performance metric for arithmetic distributions and a counter part of Lemma~\ref{lemma:composition_2stoch_procs} needs to be proved for their case.

\section{Numerical Results}\label{sect:numerical_Results}

We first simulate the model in Fig.~\ref{fig:n_hop_model} for $n=3$, i.e., a $3$-hop model with links $(0,1),(1,2)$ and $(2,3)$. We consider three different continuous probability distributions: Rayleigh with scale $\sigma=1$, Chi-Square with degree of freedom $k=1$, and Beta with shape parameters $\alpha=2$, $\beta=3$, i.e., Rayleigh(1), $\chi^2(1)$ and Beta(2,3), for inter-update times, such that they have different means. For convenience, the mean and variance of these distributions are listed in Table~\ref{table1}, along with predicted contribution of the corresponding link to age of end-user $\frac{\mathbb{E}[Y^2]}{2\mathbb{E}[Y]}$.

\begin{table}[h]
\begin{center}
\begin{tabular}{|l | l | l | l|} 
 \hline
 distribution & mean & variance &  age \\ 
 $Y$ & $\mathbb{E}[Y]$ & $var[Y]$ & $\frac{\mathbb{E}[Y^2]}{2\mathbb{E}[Y]}$ \\ 
 \hline\hline
 Rayleigh(1) & $\frac{\sigma\sqrt{\pi}}{\sqrt{2}}=1.25$  & $\frac{(4-\pi)\sigma^2}{2}=0.43$ & 0.7979 \\
 \hline
  $\chi^2(1)$ & $k=1$ & $2k=2$ & $1.5$ \\ 
 \hline
 Beta(2,3) & $\frac{\alpha}{\alpha+\beta}=\frac{2}{5}$ & $\frac{\alpha \beta(\alpha+\beta)^{-2}}{(\alpha+\beta+1)}=\frac{1}{25}$ & $0.25$ \\
 \hline
\end{tabular}
\end{center}
\caption{Distributions, means, variances, and contributions to age.}
\label{table1}
\vspace*{-0.4cm}
\end{table}

Note that Beta is supported on bounded interval [0,1], while Chi-Square and Rayleigh are supported on semi-infinite interval $[0,\infty)$. Therefore, these are all non-negative probability distributions. There are $3!=6$ ways to map the three distributions to the three links, as shown in Table~\ref{table2}. 

\begin{table}[h]
\begin{center}
\begin{tabular}{|l | l | l | l|} 
 \hline
 link~(0,1) & link~(1,2) & link~(2,3) & $\mathbb{E}[X_n(T)]$ \\ 
 \hline\hline
 Rayleigh(1) & $\chi^2(1)$ & Beta(2,3) & 2.5487 \\
 \hline
 Rayleigh(1) & Beta(2,3) & $\chi^2(1)$ & 2.5495 \\
 \hline
 $\chi^2(1)$ & Rayleigh(1) & Beta(2,3) & 2.5449 \\ 
 \hline
 $\chi^2(1)$ & Beta(2,3) & Rayleigh(1) & 2.5480 \\ 
 \hline
 Beta(2,3) & $\chi^2(1)$ & Rayleigh(1) & 2.5453 \\
 \hline
 Beta(2,3) & Rayleigh(1) & $\chi^2(1)$ & 2.5481 \\
 \hline
\end{tabular}
\end{center}
\caption{Mapping of distributions to links.}
\label{table2}
\vspace*{-0.4cm}
\end{table}

For each mapping, we simulate the network for a large duration, $T=10^3$. We take average of $X_n(T)$, which is the instantaneous age at $T=10^3$, over $2\times10^5$ iterations for each mapping to approximate $\mathbb{E}[X_n(T)]$ by the law of large numbers, which is used as a proxy for $\lim_{t \to \infty}\mathbb{E}[X_n(t)]$. Table~\ref{table2} shows that $\mathbb{E}[X_n(T)]$ numerically obtained for all mappings is close to the theoretical value of $0.7979+1.5+0.25=2.5479$ suggested by (\ref{eqn:nhop_limit_expec_age_node_n}), which is obtained by summing the link contributions of each distribution from Table~\ref{table1}.

\begin{figure}[t]
\centerline{\includegraphics[width=0.95\linewidth]{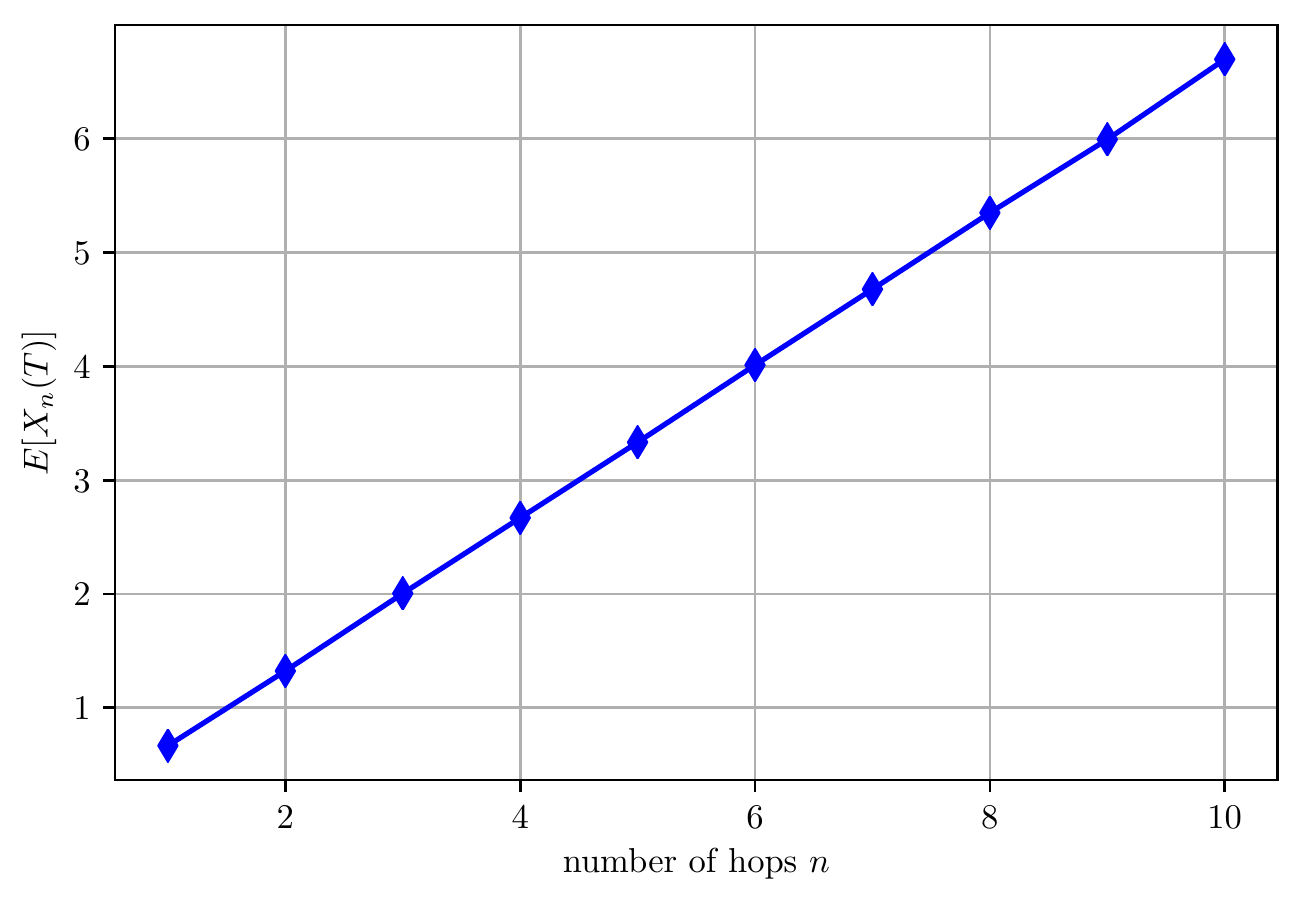}}
\vspace*{-0.1cm}
\caption{$\mathbb{E}[X_n(T)]$ in $n$-hop network with $Y^{(i,i+1)} \sim \mathcal{U}_{[0,2]}$.}
\label{fig:graph_age_vs_no_hops}
\vspace*{-0.2cm}
\end{figure}

\begin{figure}[t]
\centerline{\includegraphics[width=0.95\linewidth]{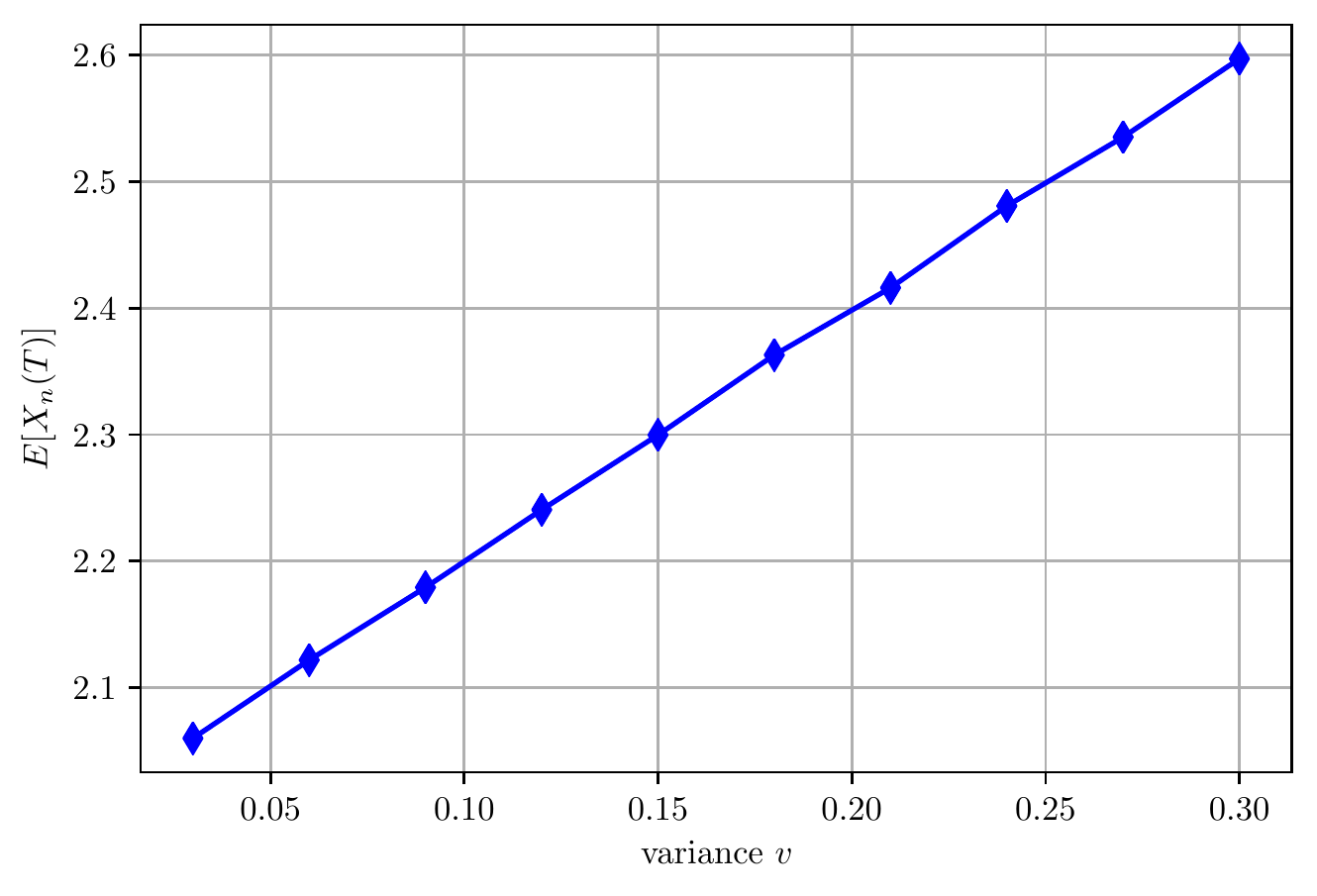}}
\vspace*{-0.1cm}
\caption{$\mathbb{E}[X_n(T)]$ in $4$-hop network with $Y^{(i,i+1)} \sim \mathcal{U}_{[1-\sqrt{3v},1+\sqrt{3v}]}$.}
\label{fig:age_vs_variance}
\vspace*{-0.4cm}
\end{figure}

Next, we simulate an $n$-hop network where update intervals of all links follow uniform distribution on the interval $[0,2]$, i.e., $Y^{(i,i+1)} \sim Y \sim \mathcal{U}_{[0,2]}$, such that $\frac{\mathbb{E}\left[Y^2\right]}{2\mathbb{E}[Y]}=\frac{2}{3}$. We plot $\mathbb{E}[X_n(T)]$ as a function of $n$ in Fig.~\ref{fig:deltas_on_timeline}. The linearity of the graph with the number of hops $n$ in Fig.~\ref{fig:deltas_on_timeline} demonstrates the additive structure of the age at the end-user as found in (\ref{eqn:nhop_limit_expec_age_node_n}). Since all links have the same distribution for inter-update times, the graph in Fig.~\ref{fig:graph_age_vs_no_hops} follows a linear equation in $n$ as $\lim_{t \to \infty}\mathbb{E}[X_n(t)]= \frac{2}{3}n$, as predicted by (\ref{eqn:nhop_limit_expec_age_node_n}).

Finally, we simulate a $4$-hop network, where inter-update times on all links follow the uniform distribution $Y^{(i,i+1)} \sim Y \sim \mathcal{U}_{[1-\sqrt{3v},1+\sqrt{3v}]}$, such that $\mathbb{E}\left[Y\right]=1$ and $var\left[Y\right]=v$. Note that for fixed mean $1$, the maximum value of $v$ is $\frac{1}{3}$ to ensure that the probability distribution has non-negative support. Fig.~\ref{fig:age_vs_variance} shows that for fixed mean, $\mathbb{E}[X_n(T)]$ increases linearly with variance $v$. Specifically, $\lim_{t \to \infty}\mathbb{E}[X_n(t)]= 4\times \frac{v+1}{2}=2v+2$ by (\ref{eqn:nhop_limit_expec_age_node_n}), as discussed in Section~\ref{sect:introduction} and Section~\ref{sect:age_treenetworks}.

\bibliographystyle{unsrt}
\bibliography{ref_priyanka}

\end{document}